\documentclass[letterpaper,12pt]{article}
\pdfoutput=1

\usepackage[left=1in,right=1in,top=1in,bottom=1in]{geometry}

\usepackage{setspace}
\usepackage[fleqn]{amsmath}
\usepackage{amssymb}
\usepackage{graphicx}
\usepackage{url}
\usepackage{verbatim}
\usepackage[title]{appendix}
\usepackage{indentfirst}
\usepackage{booktabs}
\usepackage{multirow}
\usepackage{ragged2e}
\usepackage{upgreek}
\usepackage[T1]{fontenc}
\usepackage[titles]{tocloft}
\usepackage{xspace}
\usepackage[usenames]{color}
\usepackage{ifthen}
\usepackage{cancel}
\usepackage{array}
\usepackage{tabulary}
\usepackage{authblk}
\usepackage{tikz}
\usepackage[normalem]{ulem}

\definecolor{pgreen}     {RGB}{50,162,81}
\definecolor{porange}    {RGB}{255,127,15}
\definecolor{pblue}      {RGB}{60,183,204}
\definecolor{pyellow}    {RGB}{255,217,74}
\definecolor{pteal}      {RGB}{57,115,124}
\definecolor{pauburn}    {RGB}{184,90,13}

\usepackage{hyperref}
\hypersetup{pdfborder={0 0 0},
            colorlinks=true,
            urlcolor=porange,
            linkcolor=pauburn,
            citecolor=pteal}

\usepackage[capitalize]{cleveref}

\usepackage[right, mathlines]{lineno}
\usepackage{subcaption}
\usepackage[format=plain, labelsep=period, singlelinecheck=true, skip=2pt, font=sf]{caption}
\DeclareCaptionLabelFormat{noSpace}{{#1}{#2}}
\DeclareCaptionListFormat{figList}{Figure {#2}.}
\DeclareCaptionListFormat{sFigList}{Figure S{#2}.}
\DeclareCaptionListFormat{aFigList}{Figure A{#2}.}
\usepackage{enumitem}

\newcommand{\ignore}[1]{}

\newcommand{\sub}[1]{\ensuremath{_{\textrm{#1}}}}

\newcommand{\spp}[1]{\textit{#1}}

\newcommand{\smartFigure}[4]{%
    \begin{figure}[htbp]
        \begin{center}
            \includegraphics[width=\textwidth,height=0.95\textheight,keepaspectratio]{#1}
            \captionsetup{#2}
            \caption{#3}
            \label{#4}
        \end{center}
    \end{figure}
}

\newcommand{\embedFigure}[3]{\smartFigure{#1}{listformat=figList}{#2}{#3}}

\newcommand{\embedAppendixFigure}[3]{\smartFigure{#1}{name=Figure A, labelformat=noSpace, listformat=aFigList}{#2}{#3}}
\newcommand{\given}{\ensuremath{\,|\,}\xspace}

\newcommand{\topology}[1][]{\ensuremath{T_{#1}}\xspace}
\newcommand{\data}{\ensuremath{D}\xspace}
\newcommand{\flipdata}{\ensuremath{D}\xspace}
\newcommand{\model}[1][]{\ensuremath{M_{#1}}\xspace}
\newcommand{\coinmodel}[1][]{\ensuremath{M_{#1}}\xspace}

\newcommand{\evoparameters}[1][]{\ensuremath{\theta_{#1}}\xspace}

\newcommand{\probheads}[1][]{\ensuremath{\theta_{#1}}\xspace}
\newcommand{\diff}[1]{\ensuremath{\mathrm{d}#1}}
\newcommand{\nmodels}{\ensuremath{N}\xspace}
\newcommand{\pvar}[1][]{\ensuremath{Pvar_{#1}}\xspace}
\newcommand{\new}{\ensuremath{^{\prime}}\xspace}
\newcommand{\nposteriorsamples}{\ensuremath{n}\xspace}

\usepackage[round]{natbib}

\newcommand{\jroedit}[2]{#2}
\newcommand{\jroeditb}[2]{#2}
\newcommand{\jroeditnote}[1]{}

\title{Marginal likelihoods in phylogenetics: a review of methods and applications}

\author[1]{Jamie R.\ Oaks\thanks{Corresponding author: \href{mailto:joaks@auburn.edu}{\tt joaks@auburn.edu}}}
\author[1]{Kerry A.\ Cobb}
\author[2]{Vladimir N.\ Minin}
\author[3]{Adam D.\ Leach\'{e}}
\affil[1]{Department of Biological Sciences \& Museum of Natural History, Auburn University, Auburn, Alabama 36849}
\affil[2]{Department of Statistics, University of California, Irvine, California 92697}
\affil[3]{Department of Biology \& Burke Museum of Natural History and Culture, University of Washington, Seattle, Washington 98195}

\date{\today}

\makeatletter
\let\msTitle\@title
\let\msAuthor\@author
\let\msDate\@date
\makeatother

\begin{document}


{\let\newpage\relax\maketitle}

\begin{abstract}
By providing a framework of accounting for the shared ancestry inherent to all
life,
phylogenetics is becoming the statistical foundation of biology.
The importance of model choice continues to grow as phylogenetic models
continue to increase in complexity to better capture micro and
macroevolutionary processes.
In a Bayesian framework, 
the marginal likelihood is how data update our prior beliefs about models,
which gives us an intuitive measure of comparing model fit that is grounded in
probability theory.
Given the rapid increase in the number and complexity of phylogenetic models,
methods for approximating marginal likelihoods are increasingly important.
Here we try to provide an intuitive description of marginal likelihoods and why
they are important in Bayesian model testing.
We also categorize and review methods for estimating marginal likelihoods of
phylogenetic models\jroedit{}{, highlighting several recent methods that
provide well-behaved estimates}.
\jroedit{In doing so, we use simulations to evaluate the performance of one
    such method based on approximate-Bayesian computation (ABC) and find that
    it is biased as predicted by theory.}{}
Furthermore, we review some
\jroedit{applications}{empirical studies} \jroedit{of marginal likelihoods to
phylogenetics, highlighting how they}{that demonstrate how marginal
likelihoods} can be used to learn about models of evolution from biological
data.
\jroedit{}{
We discuss promising alternatives that can complement marginal likelihoods for
Bayesian model choice, including posterior-predictive methods.
Using simulations, we find one alternative method based on approximate-Bayesian
computation (ABC) to be biased.
}
We conclude by discussing the challenges of Bayesian model choice and future
directions that promise to improve the approximation of marginal likelihoods
and Bayesian phylogenetics as a whole.

    \vspace{12pt}
\end{abstract}

\newpage

\section{Introduction}

Phylogenetics is rapidly progressing as the statistical foundation of
comparative biology, providing a framework that accounts for the shared
ancestry inherent in biological data.
Soon after phylogenetics became feasible as a likelihood-based statistical
endeavor \citep{Felsenstein1981}, models became richer to better
capture processes of biological diversification and character change.
This increasing trend in model complexity made Bayesian approaches appealing,
because they can approximate posterior distributions of rich models by
leveraging prior information and hierarchical models, where researchers can
take into account uncertainty \jroedit{of}{at} all levels in the hierarchy.

From the earliest days of Bayesian phylogenetics \citep{Rannala1996,Mau1997},
the numerical tool of choice for approximating the posterior distribution was
Markov chain Monte Carlo (MCMC).
The popularity of MCMC was due, in no small part, to avoiding the calculation
of the marginal likelihood of the model---the probability of the data under the
model, averaged, with respect to the prior, over the whole parameter space.
This marginalized measure of model fit is not easy to compute due to the large
number of parameters in phylogenetic models \jroedit{}{(including the tree
    itself)} over which the likelihood needs to be summed or integrated. 

Nonetheless, marginal likelihoods are central to model comparison in a Bayesian
framework.
\jroedit{%
If we want to compare the fit of phylogenetic models, and in the process learn
about evolutionary processes, we cannot avoid calculating marginal likelihoods.
}{%
Learning about evolutionary patterns and processes via Bayesian comparison of
phylogenetic models requires the calculation of marginal likelihoods.}
As the diversity and richness of phylogenetic models has increased, there has
been a renewed appreciation of the importance of such Bayesian model
comparison.
As a result, there has been substantial work over the last decade to develop
methods for estimating marginal likelihoods of phylogenetic models.

The goals of this review are to
(1) try to provide some intuition about what marginal likelihoods are and why
they \jroedit{are}{can be} useful,
(2) review the various methods available for approximating marginal likelihoods
of phylogenetic models,
(3) review some of the ways marginal likelihoods have been applied to learn
about evolutionary history and processes,
\jroedit{}{%
(4) highlight some alternatives to marginal likelihoods for Bayesian model
comparison,}
(5) discuss some of the challenges of Bayesian model choice, and
(6) highlight some promising avenues for advancing the field of Bayesian
phylogenetics.

\section{What are marginal likelihoods and why are they useful?}

\begin{linenomath}
A marginal likelihood is the average fit of a model to a dataset.
More specifically, it is an average over the entire parameter space of the
likelihood weighted by the prior.
For a phylogenetic model \model with parameters that include the discrete
topology (\topology) and continuous branch lengths and other parameters that
govern the evolution of the characters along the tree (together represented by
\evoparameters), the marginal likelihood can be represented as
\begin{equation}
    p(\data \given \model) =
    \sum\limits_{\topology}
    \int_{\evoparameters}
    p( \data \given \topology, \evoparameters, \model)
    p(\topology, \evoparameters \given \model)
    \diff{\evoparameters},
    \label{eq:marginalLikelihood}
\end{equation}
where \data are the data.
Each parameter of the model adds a dimension to the model, over which the
likelihood must be averaged.
The marginal likelihood is also the \jroedit{proportionality}{normalizing}
constant in the denominator of Bayes' rule that ensures the posterior is a
proper probability density that sums and integrates to one:
\begin{equation}
    p(\topology, \evoparameters \given \data, \model) = \frac{
        p(\data \given \topology, \evoparameters, \model)
        p(\topology, \evoparameters \given \model)
    }{
        p(\data \given \model)
    }.
    \label{eq:bayesRule}
\end{equation}
\end{linenomath}

Marginal likelihoods are the currency of model comparison in a Bayesian
framework.
\jroedit{The}{This differs from the} frequentist approach to model
choice\jroedit{}{, which} is based on comparing the maximum probability or density of the
data under two models either using a likelihood ratio test or some
information-theoretic criterion.
Because adding a parameter (dimension) to a model will always ensure
a maximum likelihood at least as large as without the parameter, some
penalty must be imposed when parameters are added.
\jroedit{}{%
How large this penalty should be is not easy to define, which has
led to many different possible criteria, e.g.,
the Akaike information criterion \citep[AIC;][]{Akaike1974}, 
second-order AIC \citep[AIC\sub{C};][]{Hurvich1989,Sugiura1978},
and Bayesian information criterion
\citep[BIC][]{Schwarz1978}.}

\jroedit{%
From a Bayesian perspective, we are interested in comparing the average fit of
a model, rather than the maximum.}{%
Instead of focusing on the maximum likelihood of a model, the Bayesian approach
compares the average fit of a model.}
This imposes a ``natural'' penalty for parameters, because each additional
parameter introduces a dimension that must be averaged over.
If that dimension introduces substantial parameter space with small likelihood,
and little space that improves the likelihood, it will decrease the marginal
likelihood.
Thus, unlike the maximum likelihood, adding a parameter to a model can
decrease the \emph{marginal} likelihood,
\jroedit{}{%
which ensures that more parameter-rich models are not automatically preferred.}

\begin{linenomath}
The ratio of two marginal likelihoods gives us the factor by which the
average fit of the model in the numerator is better or worse than the
model in the denominator.
This is called the Bayes factor \citep{Jeffreys1935}.
We can again leverage Bayes' rule to gain more intuition for how marginal
likelihoods and Bayes factors guide Bayesian model selection by
writing it in terms of the posterior probability of a model, \model[1],
among \nmodels candidate models:
\begin{equation}
    p(\model[1] \given \data) = \frac{
        p(\data \given \model[1])
        p(\model[1])
    }{
        \sum\limits_{i=1}^{\nmodels}
        p(\data \given \model[i])
        p(\model[i])
    }.
    \label{eq:bayesRuleModelProbability}
\end{equation}
This shows us that the posterior probability of a model is proportional to the
prior probability multiplied by the marginal likelihood of that model.
Thus, the marginal likelihood is how the data update our prior beliefs about a
model.
As a result, it is often simply referred to as ``the evidence''
\citep{MacKay2005}.
If we look at the ratio of the posterior probabilities of two models,
\begin{equation}
    \frac{
        p(\model[1] \given \data)
    }{
        p(\model[2] \given \data)
    }
    =
    \frac{
        p(\data \given \model[1])
    }{
        p(\data \given \model[2])
    }
    \times
    \frac{
        p(\model[1])
    }{
        p(\model[2])
    },
    \label{eq:modelOdds}
\end{equation}
we see that the Bayes factor is the factor by which the prior odds of a model
is multiplied to give \jroedit{use}{us} the posterior odds.
Thus, marginal likelihoods and their ratios give us intuitive measures of how
much the data ``favor'' one model over another, and these measures have natural
probabilistic interpretations.
\jroedit{}{%
However, marginal likelihoods and Bayes factors do not offer a panacea for
model choice.
As Equation~\ref{eq:marginalLikelihood} shows, weighting the average likelihood
by the prior causes marginal likelihoods to be inherently sensitive to the
prior distributions placed on the models' parameters.
To gain more intuition about what this means and how Bayesian model choice
differs from parameter estimation, let's use a simple, albeit contrived,
example of flipping a coin.}
\end{linenomath}

\subsection{A coin-flipping example}

\jroedit{%
Before we discuss methods for approximating marginal likelihoods, let's use a
simple, albeit contrived, example to help gain some intuition for marginal
likelihoods and how they differ from the posterior distribution of a model.}{}
Let's assume we are interested in the probability of a coin we have not seen
landing heads-side up when it is flipped (\probheads);
we refer to this as the rate of landing heads up to avoid confusion with other
uses of the word probability.
Our plan is to flip this coin 100 times and count the number of times it lands
heads up, which we model as a random outcome from a binomial distribution.
Before flipping, we decide to compare four models that vary in our prior
assumptions about the probability of the coin landing heads up
(Figure~\ref{fig:bayesDemo}):
We assume
\begin{enumerate}
    \item all values are equally \jroeditb{likely}{probable}
        (\coinmodel[1]: $\probheads \sim \textrm{Beta}(1, 1)$),
    \item the coin is likely weighted to land mostly ``heads'' or ``tails''
        (\coinmodel[2]: $\probheads \sim \textrm{Beta}(0.6, 0.6)$),
    \item the coin is probably fair
        (\coinmodel[3]: $\probheads \sim \textrm{Beta}(5.0, 5.0)$), and
    \item the coin is weighted to land tails side up most of time
        (\coinmodel[4]: $\probheads \sim \textrm{Beta}(1.0, 5.0)$).
\end{enumerate}
We use beta distributions to represent our prior expectations, because the
beta is a conjugate prior for the binomial likelihood function.
This allows us to obtain the posterior distribution and marginal likelihood
analytically.

\embedFigure{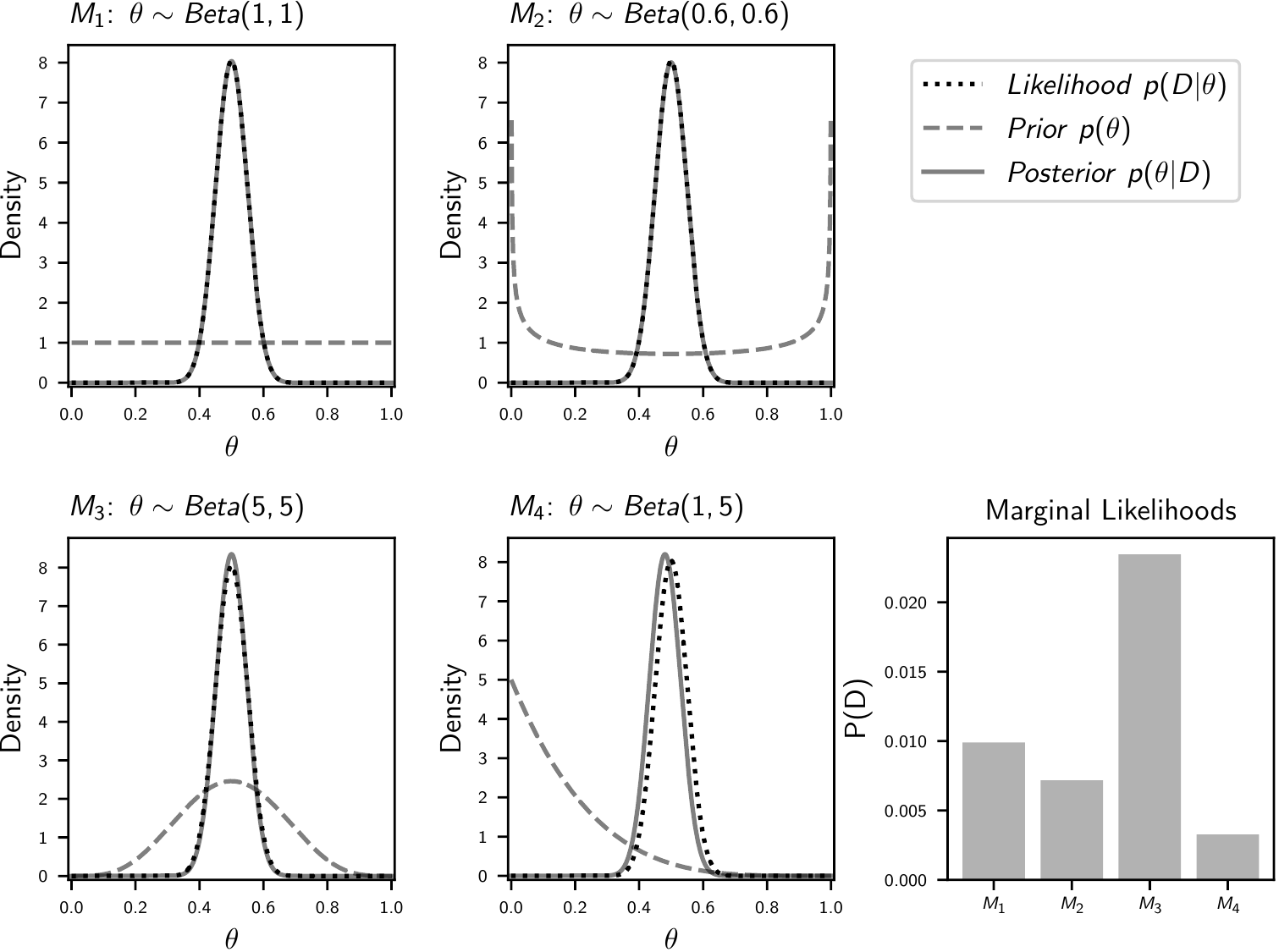}{
    An illustration of the posterior probability densities and marginal
    likelihoods of the four different prior assumptions we made in our
    coin-flipping experiment.
    The data are 50 ``heads'' out of 100 coin flips, and the parameter,
    \probheads, is the probability of the coin landing heads side up.
    The binomial likelihood density function is proportional to a
    $\textrm{Beta}(51, 51)$ and is the same across the four different beta
    priors on \probheads ($M_1$--$M_4$).
    The posterior of each model is a $\textrm{Beta}(\alpha + 50, \beta + 50)$
    distribution.
    The marginal likelihoods ($P(D)$; the average of the likelihood density
    curve weighted by the prior) of the four models are compared.
}{fig:bayesDemo}

\begin{linenomath}
After flipping the coin and observing that it landed heads side up 50 times, we
can calculate the posterior probability distribution for the rate of landing
heads up under each of our four models:
\begin{equation}
    p(\probheads \given \flipdata, \coinmodel[i]) = \frac{
        p(\flipdata \given \probheads, \coinmodel[i]) p(\probheads \given \coinmodel[i])
    }{
        p(\flipdata \given \coinmodel[i])
    }.
    \label{eq:coinBayesRule}
\end{equation}
Doing so, we see that regardless of our prior assumptions about the rate of the
coin landing heads, the posterior distribution is very similar
(Figure~\ref{fig:bayesDemo}).
This makes sense; given we observed 50 heads out of 100 flips, values for
\probheads toward zero and one are extremely unlikely, and the posterior is
dominated by the likelihood of values near 0.5.
\end{linenomath}

\begin{linenomath}
Given the posterior distribution for \probheads is very robust to our prior
assumptions, we might assume that each of our four models explain the data
similarly well.
However, to compare their ability to explain the data, we need to
average (integrate) the likelihood density function over all possible
values of \probheads, weighting by the prior:
\begin{equation}
    p(\data \given \model[i]) =
    \int_{\probheads}
    p( \data \given \probheads, \model[i])
    p(\probheads \given \model[i])
    \diff{\probheads}.
    \label{eq:coinMarginalLikelihood}
\end{equation}
Looking at the plots in Figure~\ref{fig:bayesDemo} we see that the models that
place a lot of prior weight on values of \probheads that do not explain the
data well (i.e., have small likelihood) have a much smaller marginal likelihood.
Thus, even if we have very informative data that make the posterior
distribution robust to prior assumptions, this example illustrates that the
marginal likelihood of a model can still be very sensitive to the prior
assumptions we make about the parameters.
\end{linenomath}

\jroedit{}{%
Because of this inherent sensitivity to the priors, we have to take more care
when choosing priors on the models' parameters when our goal is to compare
models versus estimating parameters.
For example, in Bayesian phylogenetics, it is commonplace to use ``uninformative''
priors, some of which are improper (i.e., they do not integrate to one).
The example above demonstrates that if we have informative data, this objective
Bayesian strategy \citep{Jeffreys1961,Berger2006} is defensible if our goal is
to infer the posterior distribution of a model;
we are hedging our bets against specifying a prior that concentrates its
probability density outside of where the true value lies, and we can rely on
the informative data to dominate the posterior.
However, this strategy is much harder to justify if our goal is to compare
marginal likelihoods among models.
First of all, models with improper priors do not have a well-defined marginal
likelihood and should not be used when comparing models \citep{Baele2013MBE}.
Second, even if diffuse priors are proper, they could potentially sink the
marginal likelihood of good models by placing excessive weight in
biologically unrealistic regions of parameter space with low likelihood.
Thus, if our goal is to leverage Bayesian model choice to learn about the
processes that gave rise to our data, a different strategy is called for.
One option is to take a more subjective Bayesian approach
\citep{Lad1996,Lindley2000,Goldstein2006} by carefully choosing prior
distributions for the models' parameters based on existing knowledge.
In the era of ``big data,'' one could also use a portion of their data to
inform the priors, and the rest of the data for inference.
Alternatively, we can use hierarchical models that allow the data to inform the
priors on the parameters \citep[e.g.,][]{Suchard2003HPM}.}

\jroedit{}{%
We have developed an interactive version of Figure~\ref{fig:bayesDemo} where
readers can vary the parameters of the coin-flip experiment and prior
assumptions to further gain intuition for marginal likelihoods
(\href{https://kerrycobb.github.io/beta-binomial-web-demo/}{https://kerrycobb.github.io/beta-binomial-web-demo/}).
It's worth noting that this pedagogical example is somewhat contrived given
that the models we are comparing are simply different priors.
Using the marginal likelihood to choose a prior is dubious, because the
``best'' prior will always be a point mass on the maximum likelihood estimate.
Nonetheless, the principles of (and differences between) Bayesian parameter
estimation and model choice that are illustrated by this example are directly
relevant to more practical Bayesian inference settings.}
Now we turn to methods for approximating the marginal likelihood of
phylogenetic models, where simple analytical solutions are
\jroedit{}{generally} not possible.
Nonetheless, the same fundamental principles apply.

\section{Methods for marginal likelihood approximation}

For all but the simplest of models, the summation and integrals in
Equation~\ref{eq:marginalLikelihood}
are analytically intractable.
This is particularly true for phylogenetic models, which have a complex
structure containing both discrete and continuous elements.
Thus, we must resort to numerical techniques to approximate the marginal
likelihood.

Perhaps the simplest numerical approximation of the marginal likelihood is to
draw samples of a model's parameters from their respective prior distributions.
This turns the intractable integral into a sum of the samples' likelihoods.
Because the prior weight of each sample is one in this case, the marginal
likelihood can be approximated by simply calculating the average likelihood of
the prior samples.
Alternatively, if we have a sample of the parameters from the posterior
distribution---like one obtained from a ``standard'' Bayesian phylogenetic
analysis via MCMC---we can again use summation to approximate the integral.
In this case, the weight of each sample is the ratio of the prior density to
the posterior density.
As a result, the sum simplifies to the harmonic mean (HM) of the likelihoods
from the posterior sample \citep{Newton1994}.
Both of these techniques can be thought of as importance-sampling integral
approximations.
\jroedit{, and suffer from}{%
Whereas both provide unbiased estimates of the marginal likelihood in theory,
they can suffer from very large Monte Carlo error due to}
the fact that the prior and posterior are often \emph{very} divergent,
with the latter usually \emph{much} more peaked than the former \jroedit{}{due
    to the strong influence of the likelihood}.
A finite sample from the prior will often yield an underestimate of the
marginal likelihood, because the region of parameter space with high likelihood
is likely to be missed.
\jroedit{Whereas}{In comparison,} a finite sample from the posterior will almost always lead to an
overestimate \citep{Lartillot2006,Xie2011,Fan2011}, because it will contain too
few samples outside of the region of high likelihood, where the prior weight
``penalizes'' the average likelihood.
\jroedit{}{%
However, \citet{Baele2016} showed that for trees with 3--6 tips and relatively
simple models, the average likelihood of a very large sample from the prior
(30--50 billion samples) can yield accurate estimates of the marginal
likelihood.
}

Recent methods developed to estimate marginal likelihoods generally fall into
two categories for dealing with the sharp contrast between the prior and
posterior that cripples the simple approaches mentioned above.
One general strategy is to turn the giant leap between the unnormalized
posterior and prior into many small steps \jroedit{}{across intermediate
    distributions};
\jroedit{}{%
methods that fall into this category require samples collected from the
intermediate distributions.}
The second strategy is to turn the giant leap between the posterior and prior
into a smaller leap between the posterior and a reference distribution that is
as similar as possible to the posterior;
\jroedit{}{%
\jroeditb{}{many}
methods in this category only require samples from the posterior
distribution.}
These approaches are not mutually exclusive (e.g., see \citet{Fan2011}), but
they serve as a useful way to categorize many of the methods available for
approximating marginal likelihoods.
In practical terms, the first strategy is computationally expensive, because
samples need to be collected from each step between the posterior and prior,
which is not normally part of a standard Bayesian phylogenetic analysis.
The second strategy \jroeditb{is}{can be} very inexpensive \jroeditb{because it}{for methods that} attempt\jroeditb{s}{} to approximate the
marginal likelihood using only the posterior samples collected from a
typical analysis.

\subsection{Approaches that bridge the prior and posterior with small steps}

\subsubsection{Path sampling (PS)}
\citet{Lartillot2006} introduced path sampling \jroedit{}{\citep{Gelman1998}}
(also called thermodynamic integration) to phylogenetics to address the problem
that the posterior is often dominated by the likelihood and very divergent from
the prior.
Rather than restrict themselves to a sample from the posterior, they collected
MCMC samples from a series of distributions between the prior and posterior.
Specifically, samples are taken from a series of power-posterior distributions,
\jroedit{}{%
$ p(\data \given \topology, \evoparameters, \model)^{\beta}
p(\topology, \evoparameters \given \model)$,}
where the likelihood is raised to a power $\beta$.
When $\beta = 1$, this is equal to the unnormalized joint posterior, which
integrates to what we want to know, the marginal likelihood.
When $\beta = 0$, this is equal to the joint prior distribution, which,
assuming we are using proper prior probability distributions, integrates to 1.
If we integrate the \jroedit{}{power} posterior expectation of the derivative
with respect to $\beta$ of the log power posterior
over the interval (0--1) with respect to $\beta$,
we get the \jroedit{}{log} ratio of the normalizing constants when $\beta$
equals 1 and 0, and since we know the constant is 1 when $\beta$ is zero, we
are left with the marginal likelihood.
\citet{Lartillot2006} approximated this integral by
summing over MCMC samples taken from a discrete number of $\beta$ values evenly
distributed between 1 and 0.

\subsubsection{\jroedit{Stepping stone (SS)}{Stepping-stone (SS) sampling}}
The stepping-stone method introduced by \citet{Xie2011}
is similar to PS in that it also uses samples from power posteriors, but the
\jroedit{motivation}{idea} is not based on approximating the integral per se, but by
the fact that we can accurately use importance sampling to approximate the
ratio of normalizing constants \jroedit{}{with respect to two pre-chosen
consecutive $\beta$ values} at each step between the posterior and prior.
Also, \citet{Xie2011} chose the values of $\beta$ for the
series of power posteriors from which to sample so that most were close to
the prior (reference) distribution, rather than evenly distributed between
0 and 1.
This is beneficial, because most of the change happens near the prior; the
likelihood begins to dominate quickly, even at small values of $\beta$.
The stepping-stone method results in more accurate estimates of the marginal
likelihood with fewer steps than PS \citep{Xie2011}.

\subsubsection{Generalized stepping stone (GSS)}
The most accurate estimator of marginal likelihoods available to date, the
generalized stepping-stone (GSS) method, combines both strategies
\jroedit{}{we are using to categorize methods by}
taking many small steps from a starting point (reference distribution) that
is much closer to the posterior than the prior \citep{Fan2011}.
\citet{Fan2011} improved upon the original stepping-stone
method by using a reference distribution that, in most cases, will be much more
similar to the posterior than the prior.
The reference distribution has the same form as the joint prior, but each
marginal prior distribution is adjusted so that its mean and variance matches
the corresponding sample mean and variance of an MCMC sample from the
posterior.
This guarantees that the support of the reference distribution will cover the
posterior.

Initially, the application of the GSS method was limited, because it required
that the topology be fixed, because there was no reference distribution across
topologies.
However, \citet{Holder2014} introduced such a distribution
on trees, allowing the GSS to approximate the fully marginalized likelihood of
phylogenetic models.
\jroedit{}{%
\citet{Baele2016} introduced additional reference distributions on trees
under coalescent models.
Furthermore,
\citet{Wu2014} and \citet{Rannala2017} showed that the GSS and PS methods
remain statistically consistent and unbiased when the topology is allowed to
vary.}

\jroedit{}{%
Based on intuition, it may seem that GSS would fail to adequately penalize
the marginal likelihood, because it would lack samples from regions of
parameter space with low likelihood (i.e., it does not use samples from the
prior).
However, importance sampling can be used to estimate the ratio of the
normalizing constant of the posterior distribution (i.e., the marginal
likelihood) to the reference distribution.
As long as the reference distribution is proper, such that its normalizing
constant is 1.0, this ratio is equal to the marginal likelihood.
As a result, any proper reference distribution that covers the same parameter
space as the posterior will work.
The closer the reference is to the posterior, the easier it is to estimate
the ratio of their normalizing constants (and thus the marginal likelihood).
In fact, at the extreme that the reference distribution matches the posterior,
we can determine the marginal likelihood exactly with only a single sample,
because the difference in their densities is solely due to the normalizing
constant of the posterior distribution
\citep{Fan2011}.}

\jroedit{}{%
All of the methods discussed below under ``Approaches that use only posterior
samples'' are based on this idea of estimating the unknown normalizing constant
of the posterior (the marginal likelihood) by "comparing" it to a reference
distribution with a known normalizing constant (or at least a known difference
in normalizing constant).
What is different about GSS is the use of samples from a series of
power-posterior distributions in between the reference and the posterior, which
make estimating the ratio of normalizing constants between each sequential pair
of distributions more accurate.}

\jroedit{}{%
The fact that PS, SS, and GSS all use samples from a series of power-posterior
distributions raises some important practical questions:
How many power-posterior distributions are sufficient, how should they be
spaced between the reference and posterior distribution, and how many MCMC
samples are needed from each?
There are no simple answers to these questions, because they will vary depending
on the data and model.
However, one general strategy that is clearly advantageous is having most of
the $\beta$ values near zero so that most of the power-posterior distributions
are similar to the reference distribution 
\citep{Lepage2007,Xie2011,Baele2016}.
Also, a practical approach to assess if the number of $\beta$ values and the
number of samples from each power posterior is sufficient is to estimate the
marginal likelihood multiple times (starting with different seeds for the
random number generator) for each model to get a measure of variance among
estimates.
It is difficult to quantify how much variance is too much, but the estimates
for a model should probably be within a log likelihood unit or two from
each other, and the ranking among models should be consistent.
It can also be useful to check how much the variance among estimates
decreases after repeating the analysis with more $\beta$ values
and/or more MCMC sampling from each step;
a large decrease in variance suggests the sampling scheme was insufficient.}

\subsubsection{Sequential Monte Carlo (SMC)}
Another approach that uses sequential importance-sampling steps is sequential
Monte Carlo (SMC), also known as particle filtering
\jroedit{}{\citep{Gordon1993,DelMoral1996,Liu1998}.}
Recently, SMC algorithms have been developed for approximating the posterior
distribution of phylogenetic trees \citep{Jordan2012,Bouchard2014,Wang2018}.
While inferring the posterior, SMC algorithms can approximate the marginal
likelihood of the model ``for free,'' by keeping a running average of the
importance-sampling weights of the trees (particles) along the way.
SMC algorithms hold a lot of promise for complementing MCMC in Bayesian
phylogenetics due to their sequential nature and ease with which the
computations can be parallelized
\citep{Jordan2012,Dinh2018,Fourment2018,Wang2018}.
See \citet{Bouchard2014} for an accessible treatment
of SMC in phylogenetics.

\jroedit{}{%
\citet{Wang2018} introduced a variant of SMC into phylogenetics that, similar
to path sampling and stepping stone, transitions from a sample from the prior
distribution to the posterior across a series of distributions where the
likelihood is raised to a power (annealing).
This approach provides an estimator of the marginal likelihood that is unbiased
from both a statistical and computational perspective.
Also, their approach maintains the full state space of the model while sampling
across the power-posterior distributions, which allows them to use standard
Metropolis-Hastings algorithms from the MCMC literature for the proposals used
during the SMC.
This should make the algorithm easier to implement in existing phylogenetic
software compared to other SMC approaches that build up the state space of the
model during the algorithm.
Under the simulation conditions they explored, \citet{Wang2018} showed that the
annealed SMC algorithm compared favorably to MCMC and SS in terms of sampling
the posterior distribution and estimating the marginal likelihood, respectively.}

\subsubsection{Nested sampling (NS)}
Recently, \citet{Maturana2017} introduced the numerical technique known as
nested sampling \jroedit{}{\citep{Skilling2006}} to Bayesian phylogenetics.
This tries to simplify the multi-dimensional integral in
Equation~\ref{eq:marginalLikelihood}
into a one-dimensional integral over the cumulative distribution function
of the likelihood.
The latter can be numerically approximated using basic quadrature methods,
essentially summing up the area of polygons under the likelihood function.
The algorithm works by starting with a random sample of parameter values
from the joint prior distribution and their associated likelihood
scores.
Sequentially, the sample with the lowest likelihood is removed and replaced by
another random sample from the prior with the constraint that its likelihood
must be larger than the removed sample.
The approximate marginal likelihood is a running sum of the likelihood of these
removed samples with appropriate weights.
Re-sampling these removed samples according to their weights yields a posterior
sample at no extra computational cost.
Initial assessment of NS suggest it performs similarly to GSS.
As with SMC, NS seems like a promising complement to MCMC for both
approximating the posterior and marginal likelihood of phylogenetic models.

\subsection{Approaches that use only posterior samples}

\subsubsection{Generalized harmonic mean (GHM)}
\citet{Gelfand1994} introduced a generalized harmonic mean
estimator that uses an arbitrary normalized reference distribution, as
opposed to the prior distribution used in the HM estimator, to weight the
samples from the posterior.
If the chosen reference distribution is more similar to the posterior than the
prior (i.e., a ``smaller leap'' as discussed above), the GHM estimator will
perform better than the HM estimator.
However, for high-dimensional phylogenetic models, choosing a suitable
reference distribution is very challenging, especially for tree topologies.
As a result, the GHM estimator has not been used for comparing phylogenetic
models.
However, recent advances on defining a reference distribution on trees
\citep{Holder2014,Baele2016} makes the GHM a tenable option in phylogenetics.

\jroedit{}{%
As discussed above, the HM estimator is unbiased in theory, but can suffer
from very large Monte Carlo error in practice.
The degree to which the GHM estimator solves this problem will depend on how
much more similar the chosen reference distribution is to the posterior
compared with the prior.
Knowing whether it is similar enough in practice will be difficult without
comparing the estimates to other unbiased methods with much smaller
Monte Carlo error (e.g., GSS, PS, or SMC).}

\subsubsection{Inflated-density ratio (IDR)}
The inflated-density ratio estimator solves the problem of choosing a
reference distribution by using a perturbation of the posterior density;
essentially the posterior is ``inflated'' from the center by a known radius
\citep{Petris2007,Arima2012,Arima2014}.
As one might expect, the radius must be chosen carefully.
The application of this method to phylogenetics has been limited by the fact
that all parameters must be unbounded; any parameters that are bounded (e.g.,
must be positive) must be re-parameterized to span the real number line,
\jroedit{}{perhaps using log transformation}.
As a result, this method cannot be applied directly to MCMC samples collected
by popular Bayesian phylogenetic software packages.
Nonetheless, the IDR estimator has recently been applied to phylogenetic models
\citep{Arima2014}, including in settings where the topology is allowed to vary
\citep{Wu2014}.
Initial applications of the IDR are very promising, demonstrating comparable
accuracy to methods that sample from power-posterior distributions while
avoiding such computation \citep{Arima2014,Wu2014}.
Currently, however, the IDR has only been used on relatively small datasets and
simple models of character evolution.
More work is necessary to determine whether the promising combination of
accuracy and computational efficiency holds for large datasets and rich models.

\subsubsection{Partition-weighted kernel (PWK)}
Recently, \citet{Wang2017} introduced the partition weighted
kernel (PWK) method of approximating marginal likelihoods.
This approach entails partitioning parameter space into regions within which
the posterior density is relatively homogeneous.
Given the complex structure of phylogenetic models, it is not obvious how this
would be done.
As of yet, this method has not been used for phylogenetic models.
However, for simulations of mixtures of bivariate normal distributions, the
PWK outperforms the IDR estimator \citep{Wang2017}.
Thus, the method holds promise if it can be adapted to phylogenetic models.

\section{Uses of marginal likelihoods}

The application of marginal likelihoods to compare phylogenetic models is
rapidly gaining popularity.
Rather than attempt to be comprehensive, below we highlight examples that
represent some of the diversity of questions being asked and the insights that
marginal likelihoods can provide about our data and the evolutionary processes
giving rise to them.

\subsection{Comparing partitioning schemes}

One of the earliest applications of marginal likelihoods in phylogenetics was
to choose among ways of assigning models of substitution to different subsets
of aligned sites.
This became important when phylogenetics moved beyond singe-locus trees to
concatenated alignments of several loci.
\citet{Mueller2004},
\citet{NylanderEtal2004}, and
\citet{Brandley2005}
used Bayes factors calculated from harmonic mean
estimates of marginal likelihoods to choose among different strategies for
partitioning aligned characters to \jroedit{}{substitution} models.
All three studies found that the model with the most subsets was strongly
preferred.
\citet{NylanderEtal2004} also showed that removing
parameters for which the data seemed to have little influence decreased the HM
estimates of the marginal likelihood, suggesting that the HM estimates might
favor over-parameterized models.
These findings could be an artefact of the tendency of the HM estimator to 
overestimate marginal likelihoods and thus underestimate the ``penalty''
associated with the prior weight of additional parameters.
However, \citet{Brown2007} showed that for simulated data,
HM estimates of Bayes factors can have a low error rate of over-partitioning an
alignment.

\citet{Fan2011} showed that, again, the HM estimator strongly
favors the most partitioned model for a four-gene alignment from cicadas (12
subsets partitioned by gene and codon position).
However, the marginal likelihoods estimated via the generalized stepping stone
method favor a much simpler model (3 subsets partitioned by codon position).
This demonstrates
\jroedit{the bias of simple importance-sampling methods}{how the HM method
fails to penalize the marginal likelihood for the weight of the prior} when
applied to finite samples from the posterior.
\jroedit{to estimate the average likelihood of
phylogenetic models.}{}
It also suggests that relatively few, well-assigned subsets can go a long way
to explain the variation in substitution rates among sites.

\jroedit{}{%
\citet{BaeleLemey2013} compared the marginal likelihoods of alternative
partitioning strategies (in combination with either strict or relaxed-clock
models) for an alignment of whole mitochondrial genomes of carnivores.
They used the harmonic mean, stabilized harmonic mean \citep{Newton1994},
path sampling, and stepping-stone estimators.
For all 41 models they evaluated, both harmonic mean estimators returned much
larger marginal likelihoods than path sampling and stepping stone, again
suggesting these estimators based solely on the posterior sample are unable to
adequately penalize the models.
They also found that by allowing the sharing of information among partitions
via hierarchical modeling \citep{Suchard2003HPM}, the model with the largest PS
and SS-estimated marginal likelihood switched from a codon model to a
nucleotide model partitioned by codon position.
This demonstrates the sensitivity of marginal likelihoods to prior assumptions.}

\subsection{Comparing models of character substitution}

\citet{Lartillot2006} used path sampling to compare
models of amino-acid substitution.
They found that the harmonic mean estimator favored the most parameter rich
model for all five datasets they explored, whereas the path-sampling estimates
favored simpler models for three of the datasets.
This again demonstrates that accurately estimated marginal likelihoods can
indeed ``penalize'' for over-parameterization of phylogenetic models.
More importantly, this work also revealed that modeling heterogeneity in amino
acid composition across sites of an alignment better explains the variation in
biological data.

\subsection{Comparing ``relaxed clock'' models}

\citet{Lepage2007} used path sampling to approximate Bayes
factors comparing various ``relaxed-clock'' phylogenetic models for
three empirical datasets.
They found that models in which the rate of substitution evolves across the tree
(autocorrelated rate models) better explain the empirical sequence alignments
they investigated than models that assume the rate of substitution on each
branch is independent (uncorrelated rate models).
This provides insight into how the rate of evolution evolves through time.

\citet{Baele2013MBE} demonstrated that modeling
among-branch rate variation with a lognormal distribution tends to explain
mammalian sequence alignments better than using an exponential distribution.
They used marginal likelihoods (PS and SS estimates) and Bayesian model
averaging to compare the fit of lognormally and exponentially distributed
\jroedit{relaxed clocks}{priors on branch-specific rates of nucleotide
    substitution (i.e., relaxed clocks)} for almost 1,000 loci from 12
mammalian species.
They found that the lognormal relaxed-clock was a better fit for almost 88\% of
the loci.
\citet{Baele2012} also used marginal likelihoods to
demonstrate the importance of using sampling dates when estimating
time-calibrated phylogenetic trees.
They used path-sampling and stepping-stone methods to estimate the marginal
likelihoods of strict and relaxed-clock models for sequence data of herpes
viruses.
They found that when the dates the viruses were sampled were provided, a strict
molecular clock was the best fit model, but when the dates were excluded,
relaxed-clock models were strongly favored.
Their findings show that using information about the ages of the tips can be
critical for accurately modeling processes of evolution and inferring
evolutionary history.

\subsection{Comparing demographic models}

\citet{Baele2012} used the path-sampling and stepping-stone
estimators for marginal likelihoods to compare the fit of various demographic
models to the HIV-1 group M data of \citet{Worobey2008},
and Methicillin-resistant \spp{Staphylococcus aureus} (MRSA)  data of \citet{Gray2011}.
They found that a flexible, nonparametric model that enforces no particular
demographic history is a better explanation of the HIV and MRSA sequence data
than exponential and logistic population growth models.
This suggests that traditional parametric growth models are not the best
predictors of viral and bacterial epidemics.

\subsection{Measuring phylogenetic information content across genomic data
    sets}

Not only can we use marginal likelihoods to learn about evolutionary models,
but we can also use them to learn important lessons about our data.
\citet{Brown2017} explored six different genomic data sets
that were collected to infer phylogenetic relationships within Amniota.
For each locus across all six data sets, they used the stepping-stone method
\citep{Xie2011} to approximate the marginal likelihood of models that included
or excluded a particular branch (bipartition) in the amniote tree.
This allowed \citet{Brown2017} to calculate, for each
gene, Bayes factors as measures of support for or against particular
relationships, some of which are uncontroversial (e.g., the monophyly of birds)
and others contentious (e.g., the placement of turtles).

\jroedit{}{%
Such use of marginal likelihoods to compare topologies, or constraints on
topologies, raises some interesting questions.
\citet{Bergsten2013} showed that using Bayes factors for topological tests can
result in strong support for a constrained topology over an unconstrained model
for reasons other than the data supporting the branch (bipartition) being
constrained.
This occurs when the data support other branches in the tree that make the
constrained branch more likely to be present just by chance, compared to a
diffuse prior on topologies.
This is not a problem with the marginal likelihoods (or their estimates),
but rather how we interpret the results of the Bayes factors;
if we want to interpret it as support for a particular relationship, we
have to be cognizant of the topology space we are summing over under
both models.
\citet{Brown2017} tried to limit the effect of this issue by constraining all
``uncontroversial'' bipartitions when they calculate the marginal likelihoods of
models with and without a particular branch, essentially enforcing an
informative prior across topologies under both models.}

Brown and Thomson's \citeyear{Brown2017} use of marginal likelihoods allowed
them to reveal a large degree of variation among loci in
support for and against relationships that was masked by the corresponding
posterior probabilities estimated by MCMC.
Furthermore, they found that a small number of loci can have a large \jroeditb{affect}{effect} on
the tree and associated posterior probabilities of branches inferred from the
combined data.
For example, they showed that including or excluding just two loci out of the
248 locus dataset of \citep{Chiari2012} resulted in a posterior probability of
1.0 in support of turtles either being sister to crocodylians or archosaurs
(crocodylians and birds), respectively.
By using marginal likelihoods of different topologies,
\citet{Brown2017} were able to identify these two loci as putative paralogs
due to their strikingly strong support for turtles being sister to
crocodylians.
This work demonstrates how marginal likelihoods can simultaneously be used as a
powerful means of controlling the quality of data in ``phylogenomics'', while
informing us about the evolutionary processes that gave rise to our data.

Furthermore, \citet{Brown2017} found that the properties
of loci commonly used as proxies for the reliability of phylogenetic signal
(rate of substitution, how ``clock-like'' the rate is, base composition
heterogeneity, amount of missing data, and alignment uncertainty) were poor
predictors of Bayes factor support for well-established amniote relationships.
This suggests these popular rules of thumb are not useful for identifying
``good'' loci for phylogenetic inference.

\subsection{Phylogenetic factor analysis}

The goal of comparative biology is to understand the relationships among a
potentially large number of phenotypic traits across organisms. 
To do so correctly, we need to account for the inherent shared ancestry
underlying all life\citep{Felsenstein1985PIC}.
A lot of progress has been made for inferring the relationship between pairs
of phenotypic traits as they evolve across a phylogeny,
but a general and efficient solution for large numbers of continuous
and discrete traits has remained elusive.
\citet{Tolkoff2017} introduced Bayesian factor analysis to
a phylogenetic framework as a potential solution.
Phylogenetic factor analysis works by modeling a small number of unobserved
(latent) factors that evolve independently across the tree, which give rise to
the large number of observed continuous and discrete phenotypic traits.
This allows correlations among traits to be estimated, without having to model
every trait as a conditionally independent process.

The question that immediately arises is, what number of factors best explains
the evolution of the observed traits?
To address this, \citet{Tolkoff2017} use path sampling to approximate the
marginal likelihood of models with different numbers of traits.
To do so, they extend the path sampling method to handle the latent variables
underlying the discrete traits by softening the thresholds that delimit the
discrete character states across the series of power posteriors.
This new approach leverages Bayesian model comparison via marginal likelihoods
to learn about the processes governing the evolution of multidimensional
phenotypes.

\subsection{Comparing phylogeographic models}

Phylogeographers are interested in explaining the genetic variation within and
among species across a landscape.
As a result, we are often interested in comparing models that include
various combinations of micro and macro-evolutionary processes and geographic
and ecological parameters.
Deriving the likelihood function for such models is often difficult and, as a
result, model choice approaches that use approximate-likelihood Bayesian
computation (ABC) are often used.

At the forefront of generalizing phylogeographic models is an approach that is
referred to as iDDC, which stands for integrating distributional, demographic,
and coalescent models \citep{Papadopoulou2016}.
This approach simulates data under various phylogeographical models upon
proxies for habitat suitability derived from species distribution models.
To choose the model \jroedit{the}{that} best explains the empirical data, this
approach \jroedit{uses}{compares} the marginal densities of the models
\jroedit{estimated via the ABC-GLM method}{%
approximated with general linear models
\citep[ABC-GLM;][]{Leuenberger2010}},
and p-values derived from these densities
\citep{He2013,Massatti2016,Bemmels2016,Knowles2017,Papadopoulou2016}.
This approach is an important step forward for bringing more biological realism
into phylogeographical models.
However, \jroedit{given}{our findings below (see section on
``approximate-likelihood approaches'' below) show}
that the marginal GLM density fitted to a truncated region of parameter space
should not be interpreted as a marginal likelihood of the full
model\jroedit{ (see above; Figure~\ref{fig:glmPerformance}),}{. Thus,}
these methods should be seen as a useful exploration of data, rather than
rigorous hypothesis tests.
Because ABC-GLM marginal densities fail to penalize parameters for their
prior weight in regions of low likelihood, these approaches will likely
be biased toward over-parameterized phylogeographical models.
Nonetheless, knowledge of this bias can help guide interpretations of results.

\subsection{Species delimitation}
Calculating the marginal probability of sequence alignments \citep{Grummer2013}
and single-nucleotide polymorphisms \citep{Leache2014} under various
multi-species coalescent models has been used to estimate species boundaries.
By comparing the marginal likelihoods of models that differ in how they assign
individual organisms to species, systematists can calculate Bayes factors to
determine how much the genetic data support different delimitations.
Using simulated data, \citet{Grummer2013} found that marginal likelihoods
calculated using path sampling and stepping-stone methods outperformed harmonic
mean estimators at identifying the true species delimitation model.
Marginal likelihoods seem better able to distinguish some species delimitation
models than others.
For example, models that lump species together or reassign samples into
different species produce larger marginal likelihood differences versus models
that split populations apart \citep{Grummer2013, Leache2014}.
Current implementations of the multi-species coalescent assume strict models of
genetic isolation, and oversplitting populations that exchange genes creates a
difficult Bayesian model comparison problem that does not include the correct
model
\jroedit{(Leach\'{e} et al.\ in prep.)}{\citep{Leache2018ME,Leache2018SSB}}. 

Species delimitation using marginal likelihoods in conjunction with Bayes
factors has some advantages over alternative approaches.
The flexibility of being able to compare non-nested models that contain
different numbers of species, or different species assignments, is one key
advantage.
The methods also integrate over gene trees, species trees, and other model
parameters, allowing the marginal likelihoods of delimitations to be compared
without conditioning on any parameters being known.
Marginal likelihoods also provide a natural way to rank competing models while
automatically accounting for model complexity \citep{Baele2012}.
Finally, it is unnecessary to assign prior probabilities to the alternative
species delimitation models being compared.
The marginal likelihood of a delimitation provides the factor by which the data
update our prior expectations, regardless of what that expectation is
(Equation~\ref{eq:bayesRuleModelProbability}).
As multi-species coalescent models continue to advance, using the marginal
likelihoods of delimitations will continue to be a powerful approach to
learning about biodiversity.

\jroedit{}{%
\section{Alternatives to marginal likelihoods for Bayesian model choice}}

\subsection{Bayesian model averaging}

\jroedit{}{%
Bayesian model averaging provides a way to avoid model choice altogether.
Rather than infer the parameter of interest (e.g., the topology) under a single
``best'' model, we can incorporate uncertainty by averaging the posterior over
alternative models.
In situations where model choice is not the primary goal, and the parameter of
interest is sensitive to which model is used, model averaging is arguably the
best solution from a Bayesian standpoint.
Nonetheless, when we jointly sample the posterior across competing models, we
can use the posterior sample for the purposes of model choice.}
\jroedit{%
An alternative to using marginal likelihoods for Bayesian model comparison is
to sample across the competing models directly.}{}
The frequency of samples from each model approximates its posterior
probability, which can be used to approximate Bayes factors among models.
\jroedit{}{%
Note, this approach is still based on marginal likelihoods---the marginal
likelihood is how the data inform the model posterior probabilities, and the
Bayes factor is simply a ratio of marginal likelihoods
(Equations \ref{eq:bayesRuleModelProbability} \& \ref{eq:modelOdds}).
However, by sampling across models, we can avoid calculating the marginal
likelihoods directly.}

Algorithms for sampling across models include reversible-jump MCMC
\citep{Green1995}, Gibbs sampling \citep{Neal2000}, Bayesian stochastic search
variable selection \citep{George1993,Kuo1998}, and approximations of reversible-jump
\citep{Jones2015}.
In fact, the first application of Bayes factors for phylogenetic model comparison was performed by \citet{Suchard2001} via reversible-jump MCMC.
This technique was also used in Bayesian tests of phylogenetic incongruence/recombination \citep{Suchard2003, Minin2005}.
In terms of selecting the correct ``relaxed-clock'' model from simulated data,
\jroedit{}{\citet{Baele2013MBE} and}
\citet{Baele2014} showed that model-averaging performed similarly to the
path-sampling and stepping-stone marginal likelihood estimators.

There are a couple of limitations for these approaches.
First, a Bayes factor that includes a model with small posterior probability
will suffer from Monte Carlo error.
For example, unless a very large sample from the posterior is collected, some
models might not be sampled at all.
\jroedit{}{%
A potential solution to this problem is adjusting the prior probabilities of
the models such that none of their posterior probabilities are very small
\citep{Carlin1995,Suchard2005}.}
Second, and perhaps more importantly, for these numerical algorithms to be able
to ``jump'' among models, the models being sampled need to be similar.
\jroedit{}{%
Whereas the first limitation is specific to using model averaging to estimate
Bayes factors, the second problem is more general.}

\jroeditb{}{%
In comparison, with estimates of marginal likelihooods in hand, we can compare
any models, regardless of how different they are in terms of parameterization
or relative probability.}
Alternatively, \jroeditb{if two highly dissimilar models need to be compared,}{}
\citet{Lartillot2006} introduced a method of using path sampling to directly
approximate the Bayes factor \jroeditb{}{between two models that can be highly
dissimilar}.
Similarly, \citet{Baele2013} extended the stepping-stone
approach of \citet{Xie2011} to do the same.
\jroeditb{%
However, if there are many models to compare, doing MCMC over power posteriors
for every pairwise comparison will quickly become computationally prohibitive;
approximating the marginal likelihood of each model would be simpler.}{}

\jroedit{}{%
\subsection{Measures of predictive performance}}

\jroedit{}{%
\begin{linenomath}
Another, albeit not an unrelated, way to compare models is based on their
predictive power, with the idea
that we should prefer the model that best predicts future data.
There are many approaches to do this, but they
are all centered around measuring the predictive power of a model using the
marginal probability of new data (\data{}\new) given our original data (\data),
\begin{equation}
    p(\data\new \given \model, \data) =
    \sum\limits_{\topology}
    \int_{\evoparameters}
    p(\data\new \given \topology, \evoparameters, \model)
    p(\topology, \evoparameters \given \model, \data)
    \diff{\evoparameters},
    \label{eq:marginalPredictiveLikelihood}
\end{equation}
which we will call the marginal posterior predictive likelihood.
This has clear parallels to the marginal likelihood (see
Equation~\ref{eq:marginalLikelihood}), with one key difference:
We condition on our knowledge of the original data, so that the average of the
likelihood of the new data is now weighted by the \emph{posterior} distribution
rather than the prior.
Thus, in situations where our data are informative and dominate the posterior
distribution under each model, the marginal posterior predictive likelihood
should be much less sensitive than the marginal likelihood to the prior
distributions used for the models' parameters.
\end{linenomath}}

\jroedit{}{%
Whether one should favor a posterior-predictive perspective or marginal
likelihoods will depend on the goals of a particular model-choice exercise
\jroeditb{}{and whether the prior is the appropriate penalty for adding
    parameters to a model}.
\jroeditb{%
Generally, marginal likelihoods will favor models that best explain
the data in hand, whereas posterior predictive approaches will
favor models that best predict future data.}{}
Regardless, posterior predictive measures of model fit are a valuable
complement to marginal likelihoods.
Methods based on the marginal posterior predictive likelihood tend to be
labeled with one of two names depending on the surrogate they use for the
``new'' data
(\data{}\new):
(1) \textbf{cross-validation methods} partition the data under study into a
training (\data) and testing (\data{}\new) dataset,
whereas
(2) \textbf{posterior-predictive methods} generate \data{}\new via simulation.}

\jroedit{}{%
\subsubsection{Cross-validation methods}}

\jroedit{}{%
\begin{linenomath}
With joint samples of parameter values from the posterior (conditional on the
training data \data), we can easily get a Monte Carlo approximation of the
marginal posterior predictive likelihood
(Equation~\ref{eq:marginalPredictiveLikelihood})
by simply taking the average probability of the testing data
across the posterior samples of parameter values:
\begin{equation}
    p(\data\new \given \model, \data) \simeq
    \frac{1}{\nposteriorsamples}
    \sum\limits_{i-1}^{\nposteriorsamples}
    p(\data\new \given \topology[i], \evoparameters[i], \model),
    \label{eq:approxMarginalPredictiveLikelihood}
\end{equation}
where \nposteriorsamples is the number of samples from the posterior under
model \model.
\citet{Lartillot2007} used this approach to show that a mixture model that
accommodates among-site heterogeneity in amino acid frequencies is a better
predictor of animal sequence alignments than standard amino-acid models.
This corroborated the findings of \citet{Lartillot2006} based on path-sampling
estimates of marginal likelihoods.
\citet{Lewis2014} introduced a leave-one-out cross-validation approach to
phylogenetics called the conditional predictive ordinates (CPO) method
\citep{Geisser1980,Gelfand1992,Chen2000}.
This method leaves one site out of the alignment to serve as the testing data
to estimate $p(\data\new \given \model, \data)$, which is equal to the
posterior harmonic mean of the site likelihood \citep{Chen2000}.
Summing the log of this value across all sites yields what is called the log
pseudomarginal likelihood (LPML).
\citet{Lewis2014} compared the estimated LPML to stepping-stone estimates of
the marginal likelihood for selecting among models that differed in how they
partitioned sites across a concatenated alignment of four genes from algae.
The LPML favored a 12-subset model (partitioned by gene and codon position)
as opposed to the 3-subset model (partitioned by codon) preferred by
marginal likelihoods.
This difference could reflect the lesser penalty against additional parameters
imposed by the weight of the posterior
(Equation~\ref{eq:marginalPredictiveLikelihood}) versus the prior
(Equation~\ref{eq:marginalLikelihood}).
\end{linenomath}}

\jroedit{}{%
\subsubsection{Posterior-predictive methods}}

\jroedit{}{%
Alternatively, we can take a different Monte Carlo approach to
Equation~\ref{eq:marginalPredictiveLikelihood} and sample from $p(\data\new
\given \model, \data)$ by simulating datasets.
For each posterior sample of the parameter values (conditional on all the data
under study) we can simply simulate a new dataset based on those parameter
values.
We can then compare the observed data (\data) to the sample of simulated
datasets (\data{}\new) from the posterior predictive distribution.
In all but the most trivial phylogenetic datasets, it is not practical to
compare the counts of site patterns directly, because there are too many
possible patterns (e.g., four raised to the power of the number of tips for DNA
data).
Thus, we have to tolerate some loss of information by summarizing the data in
some way to reduce the dimensionality.
Once a summary statistic is chosen, perhaps the simplest way to evaluate the
fit of the model is to approximate the posterior predictive p-value by finding
the percentile of the statistic from the observed data out of the values of the
statistic calculated from the simulated datasets \citep{Rubin1984,Gelfand1992}.
\citet{Bollback2002} explored this approach for phylogenetic models using
simulated data,
and found that a simple JC69 model \citep{JC1969} was often rejected for data
simulated under more complex K2P \citep{K2P} and GTR \citep{Tavare1997} models.
\citet{Lartillot2007} also used this approach to corroborate their findings
based on marginal likelihoods \citep{Lartillot2006} and cross validation that
allowing among-site variation in amino acid composition (i.e., the CAT model)
leads to a better fit.}

\jroedit{}{%
One drawback of the posterior predictive p-value is that it rewards models with
large posterior predictive variance \citep{Lewis2014}.
In other words, a model that produces a broad enough distribution of datasets
can avoid the observed data falling into one of the tails.
The method of \citet{GelfandGosh1998} (GG) attempts to solve this problem by
balancing the tradeoff between posterior predictive variance and
goodness-of-fit.
\citet{Lewis2014} introduced the GG method into
phylogenetics and compared it to cross-validation (LPML) and stepping-stone
estimates of marginal likelihoods for selecting among models that differed
in how they partitioned the sites of a four-gene alignment of algae.
Similar to LPML, the GG method preferred the model with most subsets (12;
partitioned by gene and codon position),
in contrast to the marginal likelihood estimates, which favored the model
partitioned by codon position (three subsets).
Again, this difference could be due to the lesser penalty against parameters
imposed by the weight of the posterior
(Equation~\ref{eq:marginalPredictiveLikelihood}) versus the prior
(Equation~\ref{eq:marginalLikelihood}).}

\subsection{Approximate-likelihood approaches}

Approximate-likelihood Bayesian computation (ABC) approaches
\citep{Tavare1997,Beaumont2002} have become popular in situations where it is
not possible (or undesirable) to derive and compute the likelihood function of
a model.
The basic idea is simple: by generating simulations under the model, the
fraction of times that we generate a simulated dataset that matches the
observed data is a Monte Carlo approximation of the likelihood.
Because simulating the observed data exactly is often not possible (or extremely
unlikely), simulations ``close enough'' to the observed data are
counted, and usually a set of insufficient summary statistics are used in place
of the data.
Whether a simulated dataset is ``close enough'' to count is formalized as
whether or not it falls within a zone of tolerance around the empirical data.

This simple approach assumes the likelihood within the zone of tolerance is
\jroedit{uniform}{constant}.
However, this zone usually needs to be quite large for computational
tractability, so this assumption does not hold.
Leuenberger and Wegmann \citep{Leuenberger2010} proposed fitting a general
linear model (GLM) to approximate the likelihood within the zone of tolerance.
With the GLM in hand, the marginal likelihood of the model can simply be
approximated by the marginal density of the GLM.

The accuracy of this estimator has not been assessed.
However, there are good theoretical reasons to be skeptical of its accuracy.
Because the GLM is only fit within the zone of tolerance (also called the
``truncated prior''), it cannot account for the weight of the prior on the
marginal likelihood outside of this region.
Whereas the posterior distribution usually is not strongly influenced by
regions of parameter space with low likelihood, the marginal likelihood very
much is.
By not accounting for prior weight in regions of parameter space outside the
zone of tolerance, where the likelihood is low, we predict this method will not
properly penalize models and tend to favor models with more parameters.
\jroedit{%
This is analogous with how the harmonic mean estimator tends to overestimate
the marginal likelihood due to having very few samples outside of the region of
high likelihood.}{}

To test this prediction, we assessed the behavior of the ABC-GLM method on
100 datasets simulated under the simplest possible phylogenetic model: two DNA
sequences separated by a single branch along which the sequence evolved under a
Jukes-Cantor model of nucleotide substitution \citep{JC1969}.
The simulated sequences were 10,000 nucleotides long, and the prior on the only
parameter in the model, the length of the branch, was a uniform distribution
from 0.0001 to 0.1 substitutions per site.
For such a simple model, we used quadrature integration to calculate the
marginal likelihood for each simulated alignment of two sequences.
Integration using 1,000 and 10,000 steps and rectangular and trapezoidal
quadrature rules all yielded identical values for the log marginal likelihood
to at least five decimal places for all 100 simulated data sets, providing a
very precise proxy for the true values.
We used a sufficient summary statistic, the proportion of variable sites, for
ABC analyses.
However, the ABC-GLM and quadrature marginal likelihoods are not directly
comparable, because the marginal probability of the proportion of variable
sites versus the site pattern counts will be on different scales that are data
set dependent.
So, we compare the ratio of marginal likelihoods (i.e., Bayes factors)
comparing the correct branch-length model
[branch length $\sim$ uniform(0.0001, 0.1)]
to a model with a prior approximately twice as broad
[branch length $\sim$ uniform(0.0001, 0.2)].
\jroedit{}{%
As we noted in our coin-flipping example, using marginal likelihoods to compare
priors is dubious, and we do not advocate selecting priors in this way.
However, in this case, comparing the marginal likelihood under these two priors
is useful, because it allows us to directly test our prediction that the
ABC-GLM method will not be able to correctly penalize the marginal likelihood
for the additional parameter space under the broader prior.}

This very simple model is a good test of the ABC-GLM marginal likelihood
estimator for several reasons.
The use of a sufficient statistic for a finite, one-dimensional model makes ABC
nearly equivalent to a full-likelihood Bayesian method
(Figure~A\ref{fig:branchLengthEstimates}).
Thus, this is a ``best-case scenario'' for the ABC-GLM approach.
Also, we can use quadrature integration for very good proxies for the true
Bayes factors.
Lastly, the simple scenario gives us some analytical expectations for the
behavior of ABC-GLM.
If it cannot penalize the marginal likelihood for the additional branch length
space in the model with the broader prior, the Bayes factor should be off by a
factor of approximately 2, or more precisely $(0.2-0.0001) / (0.1-0.0001)$.
As shown in Figure~\ref{fig:glmPerformance}, this is exactly what we find.
This confirms our prediction that the ABC-GLM approach cannot average over
regions of parameter space with low likelihood and thus will be biased toward
favoring models with more parameter\jroedit{ space}{s}.
Given that the GLM approximation of the likelihood is only fit within a subset
of parameter space with high likelihood, which is usually a \emph{very} small
region of a model, the marginal of the GLM should not be considered a marginal
likelihood of the model.
We want to emphasize that our findings in no way detract from the usefulness of
ABC-GLM for parameter estimation.

Full details of these analyses, which were all designed atop the DendroPy
phylogenetic API (version 4.3.0 commit 72ce015) \citep{Sukumaran2010}, can be
found in
\jroedit{the supplementary materials}{Appendix~\ref{appendix:methods}},
and all of the code to replicate our
results is freely available at
\url{https://github.com/phyletica/abc-glm-marginal-test}.

\embedFigure{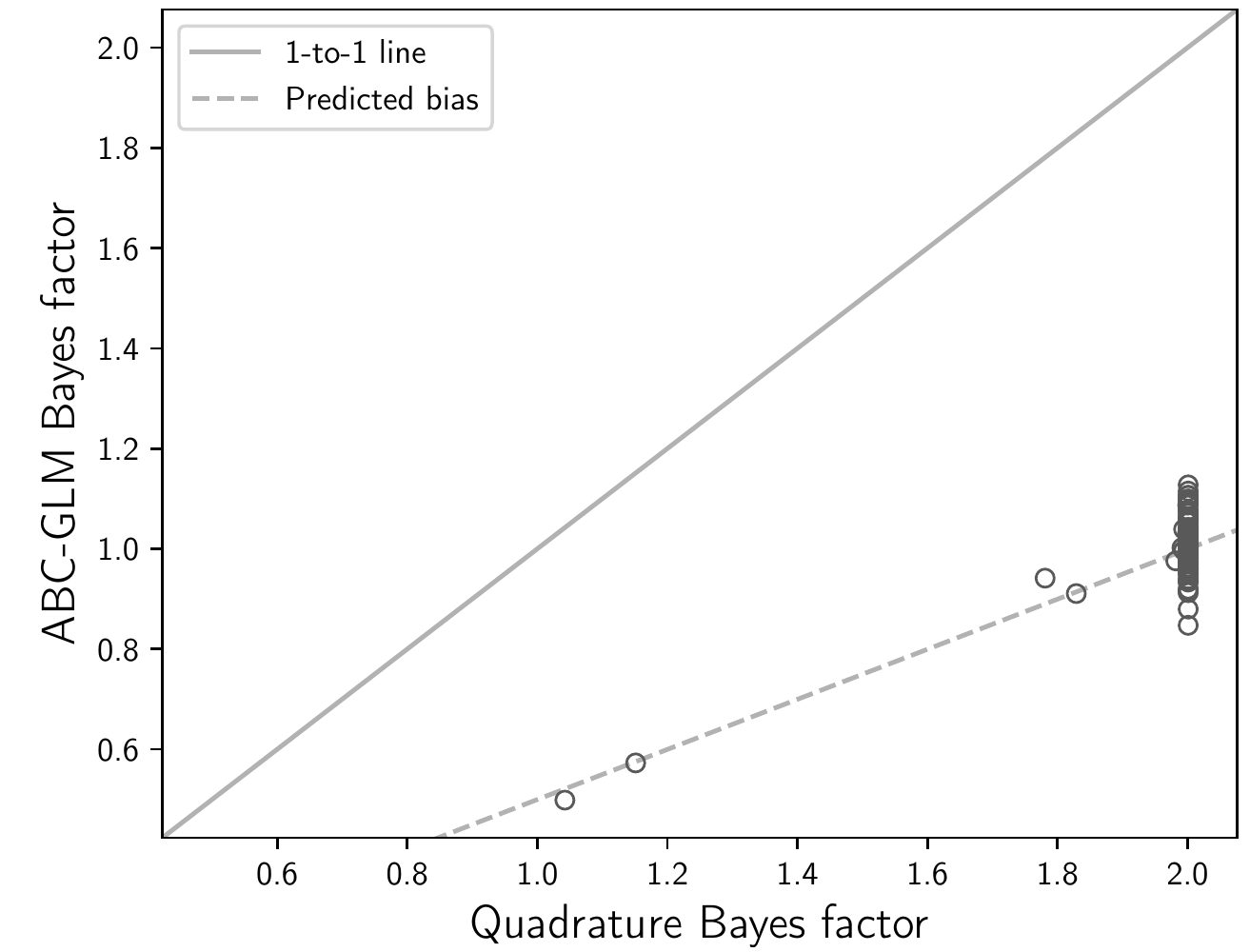}{
    A comparison of the approximate-likelihood Bayesian computation general
    linear model (ABC-GLM) estimator of the marginal likelihood
    \citep{Leuenberger2010} to quadrature integration approximations
    \citep{Xie2011} for 100 simulated datasets.
    We compared the ratio of the marginal likelihood (Bayes factor) comparing
    the correct branch-length model
    [branch length $\sim$ uniform(0.0001, 0.1)]
    to a model with a broader prior on the branch length
    [branch length $\sim$ uniform(0.0001, 0.2)].
    The solid line represents perfect performance of the ABC-GLM estimator
    (i.e., matching the ``true'' value of the Bayes factor).
    The dashed line represents the expected Bayes factor when failing to
    penalize for the extra parameter space (branch length 0.1 to 0.2) with
    essentially zero likelihood.
    Quadrature integration with 1,000 and 10,000 steps using the rectangular
    and trapezoidal rule produced identical values of log marginal likelihoods
    to at least five decimal places for all 100 simulated datasets.
}{fig:glmPerformance}

\section{Discussion}

\subsection{Promising future directions}

As Bayesian phylogenetics continues to explore more complex models of
evolution, and datasets continue to get larger, accurate and efficient methods
of estimating marginal likelihoods will become increasingly important.
Thanks to substantial work in recent years, robust methods have been developed,
such as the generalized stepping-stone approach \citep{Fan2011}.
However, these methods are computationally demanding as they have to sample
likelihoods across a series of power-posterior distributions that are not
useful for parameter estimation.
Recent work has introduced promising methods to estimate marginal likelihoods
solely from samples from the posterior distribution.
However, these methods remain difficult to apply to phylogenetic models, and
their performance on rich models and large datasets remains to be explored.

Promising avenues for future research on methods for estimating marginal
likelihoods of phylogenetic models include continued work on reference
distributions that are as similar to the posterior as possible, but easy to
formulate and use.
This would improve the performance and applicability of the GSS and derivations
of the GHM approach.
Currently, the most promising method that works solely from a posterior
sample is IDR.
Making this method easier to apply to phylogenetic models and implementing
it in popular Bayesian phylogenetic software packages,
like
RevBayes \citep{Hohna2016}
and
BEAST \citep{Suchard2018,Bouckaert2014}
would be very useful, though nontrivial.

Furthermore, nested sampling and sequential Monte Carlo are exciting numerical
approaches to Bayesian phylogenetics.
These methods essentially use the same amount of computation to both sample
from the posterior distribution of phylogenetic models and provide an
approximation of the marginal likelihood.
Both approaches are relatively new to phylogenetics, but hold a lot of promise
for Bayesian phylogenetics generally and model comparison via marginal
likelihoods specifically.

\jroeditnote{Section titled ``The state of ABC approaches to Bayesian model choice'' was removed}

\subsection{A fundamental challenge of Bayesian model choice}

While the computational challenges to approximating marginal likelihoods
are very real and will provide fertile ground for future research,
it is often easy to forget about a fundamental challenge of Bayesian model
choice.
This challenge becomes apparent when we reflect on the differences between
Bayesian model choice and parameter estimation.
The posterior distribution of a model, and associated parameter estimates, are
informed by the likelihood function (Equation~\ref{eq:bayesRule}),
whereas the posterior probability of that model is informed by the
\emph{marginal} likelihood
(Equation~\ref{eq:bayesRuleModelProbability}).
When we have informative data, the posterior distribution is dominated by the
likelihood, and as a result our parameter estimates are often robust to prior
assumptions we make about the parameters.
However, when comparing models, we need to assess their overall ability to
predict the data, which entails averaging over the entire parameter space of
the model, not just the regions of high likelihood.
As a result, marginal likelihoods and associated model choices can be very
sensitive to priors on the \emph{parameters} of each model, even when the data
are very informative (Figure~\ref{fig:bayesDemo}).
This sensitivity to prior assumptions about parameters is inherent to Bayesian
model choice
\jroedit{}{%
based on marginal likelihoods (i.e., Bayes factors and Bayesian model
averaging).
However, other Bayesian model selection approaches, such as cross-validation
and posterior-predictive methods, will be less sensitive to prior assumptions.}
\jroedit{Accordingly}{Regardless}, the results of any application of Bayesian
model selection should be accompanied by an assessment of the sensitivity of
those results to the priors placed on the models' parameters.

\subsection{Conclusions}

Marginal likelihoods are intuitive measures of model fit that are grounded in
probability theory.
As a result, they provide us with a coherent way of gaining a better
understanding about how evolution proceeds as we accrue biological data.
We highlighted how marginal likelihoods of phylogenetic models can be used to
learn about evolutionary processes and how our data inform our models.
Because shared ancestry is a fundamental property of life, the use of marginal
likelihoods of phylogenetic models promises to continue to advance biology.

\section{Funding}
This work was supported by the National Science Foundation (grant numbers DBI
1308885 and DEB 1656004 to JRO).

\section{Acknowledgments}
We thank Mark Holder for helpful discussions about comparing approximate and
full marginal likelihoods.
We also thank \jroedit{}{Ziheng Yang and} members of the Phyletica Lab (the
phyleticians) for helpful comments that improved an early draft of this paper.
\jroedit{}{
We are grateful to Guy Baele, Nicolas Lartillot, Paul Lewis, two anonymous
referees, and Associate Editor, Olivier Gascuel, for constructive reviews that
greatly improved this work.
}
The computational work was made possible by the Auburn University (AU) Hopper
Cluster supported by the AU Office of Information Technology.
This paper is contribution number 880 of the Auburn University
Museum of Natural History.

\begin{appendices}
\setcounter{figure}{0}
\section{Methods for assessing performance of ABC-GLM estimator}
\label{appendix:methods}

We set up a simple scenario for assessing the performance of the method for
estimating marginal likelihoods based on approximating the likelihood function
with a general linear model (GLM) fitted to posterior samples collected via
approximate-likelihoood Bayesian computation (ABC) \citep{Leuenberger2010};
hereforth referred to as ABC-GLM.
The scenario is a DNA sequence, 10,000-nucleotides in length, that evolves
along a branch according to a Jukes-Cantor continuous-time Markov chain (CTMC)
model of nucleotide substitution \citep{JC1969}.
Because the Jukes-Cantor model forces the relative rates of change among the
four nucleotides and the equilibrium nucleotide frequencies to be equal, there
is only a single parameter in the model, the length of the branch, and the
direction of evolution along the branch does not matter.

\subsection{Simulating data sets}
We simulated 100 data sets under this model by
\begin{enumerate}
    \item drawing 10,000 nucleotides of the ``ancestral'' sequence from their
        equilibrium frequencies ($\frac{1}{4}$), 
    \item drawing a branch length $\sim$ uniform(0.0001, 0.1), and
    \item evolving the sequence along the branch according to the Jukes-Cantor
        CTMC model to get the ``descendant'' sequence.
\end{enumerate}
This was done using the DendroPy phylogenetic API (version 4.3.0 commit
72ce015) \citep{Sukumaran2010}.

\subsection{Calculating ``true'' Bayes factors}
For each data set, we used quadrature approaches to approximate the marginal
likelihood by integrating the posterior density over the branch length prior.
We did this for two models:
\begin{enumerate}
    \item the correct model [branch length $\sim$ uniform(0.0001, 0.1)], and
    \item a model with a branch length prior slightly more than twice as broad
        [branch length $\sim$ uniform(0.0001, 0.2)], which we refer to as the
        ``vague model''.
\end{enumerate}
For both models and for each dataset we used the rectangular and trapezoidal
quadrature rules with 1,000 and 10,000 steps (i.e., four approximations of the
marginal likelihood for each data set under each model).
Across all 100 data sets and both models, all four approximations were
identical to at least five decimal places.
For each data set, we calculated the log Bayes factor comparing the correct
model to the vague model.

\subsection{Approximate-likelihood Bayesian computation}
To collect an approximate posterior sample from the correct model for a data
set, we first calculated the proportion of variable sites (\pvar) between the
two sequences.
Next, we simulated 50,000 datasets under the correct model, calculated \pvar
for each of them, and retained the 1,000 samples with the values of \pvar
closest to that calculated from the data.
Lastly, we used ABCtoolbox version 1.1 \cite{ABCtoolbox} to fit a GLM to the
retained samples and calculate the marginal density of the GLM, using a
bandwidth of 0.002.
We did the same to obtain an ABC-GLM estimate of the marginal density
for the vague model with two differences:
(1) we drew the branch length for each prior sample from the vague prior
[branch length $\sim$ uniform(0.0001, 0.2)], and
(2) to maintain the same expected tolerance under both models, we simulated
100,000 datasets under the vague model (retaining the 1,000 samples closest to
the \pvar of the data).

For each data set, we calculated the log Bayes factor from the GLM marginal
densities of the correct and vague model, and compared the ABC-GLM-estimated
Bayes factor to the ``true'' Bayes factor calculated via quadrature
integration (Figure~\ref{fig:glmPerformance}).

\subsection{Full-likelihood Markov chain Monte Carlo analyses}
One goal of the simplicity of the above model is that the additional
approximation of the ABC approach would be limited.
All numerical Bayesian analyses, based on full or approximate likelihoods,
suffer from Monte Carlo error associated with approximating the posterior with
a finite number of samples.
Approximate-likelihood methods usually suffer from two additional sources of
approximation:
(1) the full data are replaced with insufficient summary statistics, and
(2) samples are retained that do not exactly match the data or summary
statistics (i.e., the ``tolerance'' of ABC).
In our analyses described above, we avoided the former source of error by
using a sufficient statistic.
We hoped to minimize the latter source of error by evaluating many samples from
a one-dimensional model with finite bounds; we also kept this source of error
approximately equal for both models by sampling in proportion to the width of
the model.

To verify that the error introduced by the tolerance of the ABC analyses was
minimal, we compared the branch length estimates to those estimated by
full-likelihood Markov chain Monte Carlo (MCMC).
For each data set, under both models, we ran a chain for 10,000 generations,
sampling every 10 generations.
All chains appeared to reach stationarity by the first sample (10th
generation).
We plotted the branch length estimated via ABC-GLM and MCMC under both
the true and vague models against the true branch lengths.
The results of all four analyses across all 100 data sets are almost
indistinguishable (Figure~A\ref{fig:branchLengthEstimates}), confirming that
the approximation introduced by the tolerance is very minimal.
Our ABC-GLM analyses are essentially equivalent to full-likelihood Bayesian
analyses, creating a ``best-case scenario'' for evaluating the marginal
likelihood estimates of the ABC-GLM method.

\embedAppendixFigure{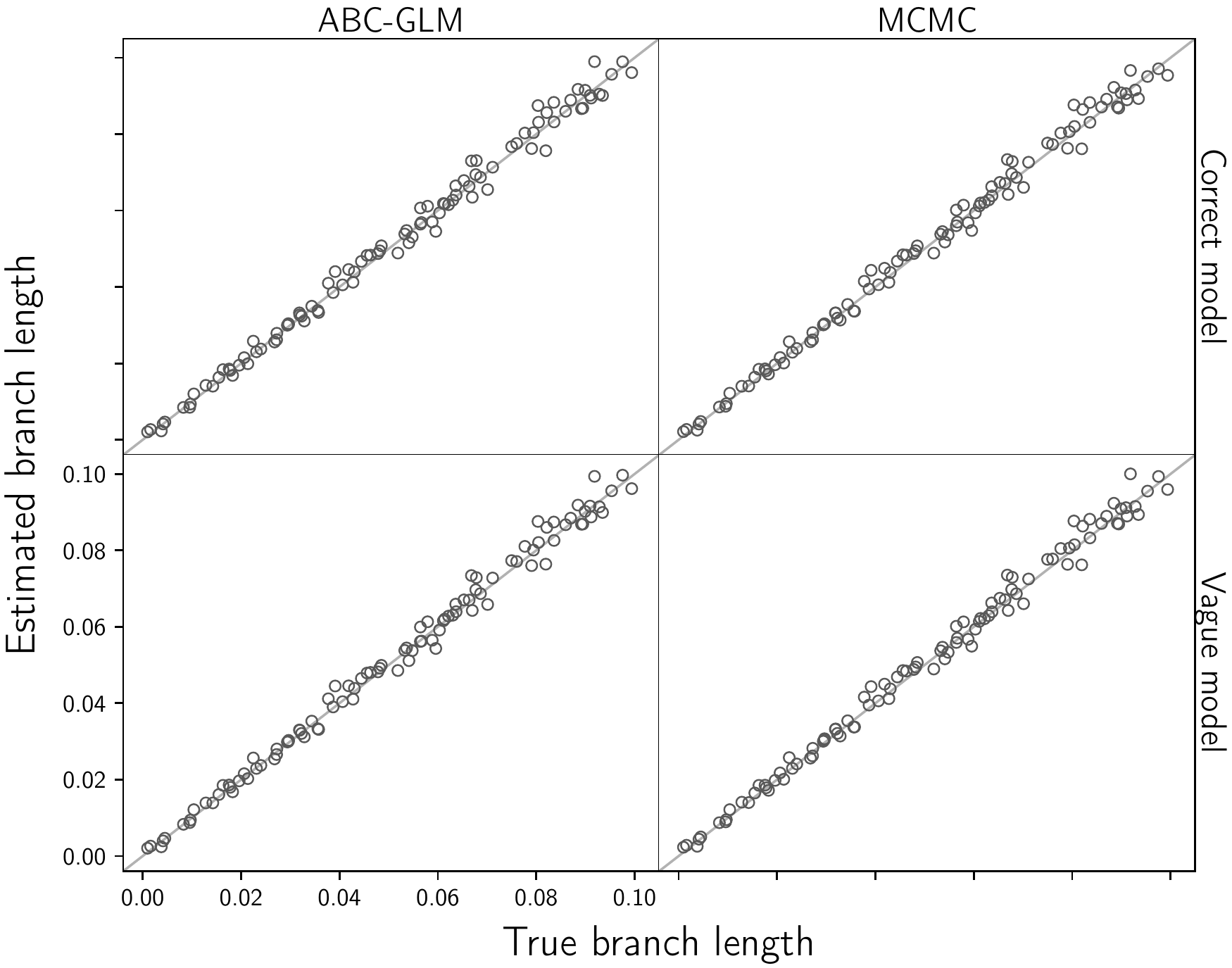}{
    A comparison of the true branch length separating each pair of simulated
    sequences to the branch length estimated by ABC-GLM and full-likelihood
    MCMC under the correct branch-length model (branch length $\sim$ uniform(0.0001,
    0.1)) and the vague model (branch length $\sim$ uniform(0.0001, 0.1)).
}{fig:branchLengthEstimates}

\subsection{Reproducibility}
All of the code to replicate our results is freely available at
\href{https://github.com/phyletica/abc-glm-marginal-test}{https://github.com/phyletica/abc-glm-marginal-test}.

\end{appendices}

\bibliographystyle{sysbio}

\end{document}